 \documentclass[final,1p,times,authoryear]{elsarticle}

\usepackage{amssymb}
\usepackage{amsmath}

\usepackage{graphicx}
\usepackage{float}
\usepackage{subfig}
\usepackage{dcolumn}
\usepackage{bm}
\usepackage{amsmath}
\usepackage{accents}
\usepackage{xr}
\usepackage{xr-hyper}
\usepackage[colorlinks=true, allcolors=blue]{hyperref}
\usepackage{nameref}
\usepackage{tikz}
\usepackage{siunitx} 
\usepackage{natbib}
\usepackage{widetext}
\begin{document}

\begin{frontmatter}



\title{The interactions between two drops floating on a partially miscible liquid pool} 


\author[UCAS]{Yuan Gao} 
 \author[UCAS,CAS]{Yanshen Li\corref{cor1}}
 \affiliation[UCAS]{organization={School of Engineering Science},
             addressline={University of Chinese Academy of Sciences },
             city={Beijing},
            postcode={101408},
             country={PR China}}

 \affiliation[CAS]{organization={State Key Laboratory of Nonlinear Mechanics},
            addressline={Institute of Mechanics},
             city={Beijing},
             postcode={100190},
            country={PR China}}
\cortext[cor1]{Email address for correspondence:liyanshen@ucas.ac.cn}

\begin{abstract}
The interaction of drops floating on liquid surfaces is important for many natural processes and industrial applications. In many of the cases, the system is multicomponent, leading to Marangoni flows on the surface. Here we investigate the competing effect of the attractive ``Cheerios effect'' and the repulsive solutal Marangoni flow by observing the behaviors of two identical oil drops floating on partially miscible pool made of ethanol-water mixtures. Three typical behaviors are found: Repel, Coalesce and Rebound, in which the drops repel each other, attract each other and then coalesce, and attract and rebound upon contact. A scaling theory based on the two competing forces is developed to distinguish the repulsive and attractive behaviors of the drops. For the transition from Coalesce to Rebound, a lubrication layer is found to form when the immersed lower halves of the drops are more than half a sphere, which prevents the drops from coalescing. 
\end{abstract}




\end{frontmatter}



\section{\label{Introduction}Introduction}
The interaction of droplets on liquid surfaces has attracted lots of interest not only due to their applications in petroleum industry \citep{chen1990experimental, Oil_Spills} and environmental protection \citep{kornilios1998pelagic, li2005waste}, but also due to scientific interests because intriguing phenomena like droplet dancing \citep{cejkova2019dancing}, droplet oscillations \citep{chen2017spontaneous} and even faceted droplets \citep{li2023oilonwater} can emerge. These intriguing phenomena are caused by Marangoni flows as a consequence of the solute gradient developed on the liquid surface. Marangoni flows have been known to trigger the self-propulsion of drops on liquid surfaces \citep{nagai2005mode,chen2009self, pimienta2014selfpropulsion, tanaka2015spontaneous}. When two drops are present, the Marangoni flow may introduce repulsive forces between them. On the other hand, when the drops are not too small, their gravity may lead to the deformation of the liquid surface around them, thus leading to attractive forces, which is often called the ``Cheerios effect'' \citep{Nicolson_1949,Vella2004TheE}. Though the ``Cheerios effect'' and the effect of Marangoni flows on drops have been studied separately, the combined effects of these two factors have not been investigated throughly.

Here we systematically investigate the effects of these two factors by observing the behaviors of two identical oil drops floating a partially miscible liquid pool made of ethanol-water mixtures. Dissolution of the drop lowers the surface tension of the mixture, leading to an outward Marangoni flow around the drop. The Marangoni flow causes repulsive forces between the two drops while the ``Cheerios effect'' causes attractive forces. The drop volume and the ethanol concentration of the pool are varied to span the two-dimensional parameter space. Three typical behaviors are observed: Repel, Coalesce and Rebound, in which the drops repel each other, attract each other and coalesce, and attract each other but rebound upon contact, respectively. A scaling theory is developed based on the competition between the two forces, which can separate the repulsive behaviors (Repel) and attractive behaviors (Coalesce and Rebound) well. Later, a lubrication layer is observed upon contact when the immersed part of the drop is more than half of a sphere, which causes the transition from Coalesce to Rebound.

\section{\label{sec:ecEXPERIMENT}Experimental procedure and methods}

\begin{figure}[h]
\centering
\includegraphics [width=0.5\textwidth]{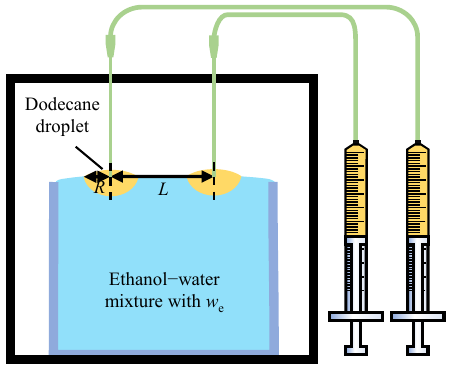}
\caption{Sketch of the experimental setup. A glass container is filled with ethanol-water mixture and two dodecane drops of the same volume are deposited onto the surface of the mixture via two needles. The container is put in a lager container made of acrylic glass to prevent the preferential evaporation of ethanol and subsequent interfacial flows. The behaviors of the drops are monitored from above, from which the drop radius $R$ and the central distance $L$ between the drops are measured. The ethanol weight fraction of the liquid pool is $w_\mathrm{e}$.
} 
\label{Fig:The experiment setup}
\end{figure}

In the experiment, dodecane (Macklin, $\geq\SI{99.0}{\%}$) is used as the drop phase and ethanol-water mixtures are used as the liquid pool. Deionized water (\SI{18}{M\Omega\cdot\cm}) 
and ethanol (Innochem, \SI{99.7}{\%}) are used to make the mixtures. Because mixing ethanol and water is an exothermic process, the ethanol-water mixtures are always prepared at least one day in advance for them to reach thermal equilibrium with the surroundings. Since the density of dodecane $\rho=\SI{749.5}{\kg/m^3}$ is smaller than that of ethanol, dodecane drops will always float on the ethanol-water mixture. The mixture is put in a glass container with size $50\times 50 \times \SI{40}{\mm^3}$. To prevent the drop from being attracted to the side walls of the glass container, the shape of the meniscus at the container wall -- either pulled up or pushed down -- is controlled by setting the depth of the mixture slightly lower or higher than $\SI{40}{\mm}$. In addition, to prevent the preferential evaporation of ethanol and the subsequent interfacial flows, the liquid pool is put in a larger container covered with a lid. They are both made of acrylic glass and the container's inner size is $65\times 65 \times \SI{65}{\mm^3}$. On the lid, two holes with diameter \SI{0.5}{\mm} and \SI{15}{\mm} apart are made to facilitate drop deposition, thus the initial central distance between the two drops is $L_0=\SI{15}{\mm}$. See Fig.\ref{Fig:The experiment setup} for the sketch of the experimental setup. Ten minutes after covering the lid, two dodecane drops of the same volume are deposited at the same time via two needles through the holes. The two needles are connected to two syringes which are mounted on one syringe pump with two racks (ZhongXinQiHeng, Cchippump-2, China). To record the motion and shape of the drops, top and side views are captured by two cameras (Nikon D850) each connected to a long working distance lens system (Thorlabs, MVL12X12Z plus 0.25X or 0.5X lens attachment). 

To fully explore the parameter space, the drop volume $V$ is varied between \SI{1}{\ul} to \SI{370}{\ul} and the ethanol weight fraction $w_\mathrm{e}$ of the mixture is varied between \SI{10}{wt\%} and \SI{90}{wt\%}. The surface tension of the ethanol mixture is obtained from the literature \citep{lide2004crc} and the surface tension of the dodecane saturated ethanol-water mixture is measured by the pendant drop method using a goniometer (KR\"{U}SS, DSA25S, Germany). The dodecane saturated ethanol-water mixtures are prepared in the following way: The ethanol-water mixture at the desired ethanol weight fraction $w_\mathrm{e}$ is first prepared with a total mass of $\approx\SI{100}{\g}$. A drop of dodecane is added to the mixture and shaken rigorously. This process is repeated until phase separation happens \citep{Li_Chong_Bazyar_Lammertink_Lohse_2021}. To make sure the mixture has reached thermal equilibrium with the environment, the saturated mixture was also made one day in advance. 

To obtain the flow dynamics, Particle Image Velocimetry (PIV) measurements were performed by adding tracer particles (Dantec Dynamics, Polyamide Seeding Particles, Denmark) \SI{5}{\um} in diameter to the ethanol-water mixture beforehand. The particles were found to follow the flow faithfully, see Supplementary Material for details.

\section{\label{sec:drop behavior}Experimental results}
\begin{figure*}[!h]
\centering
\includegraphics {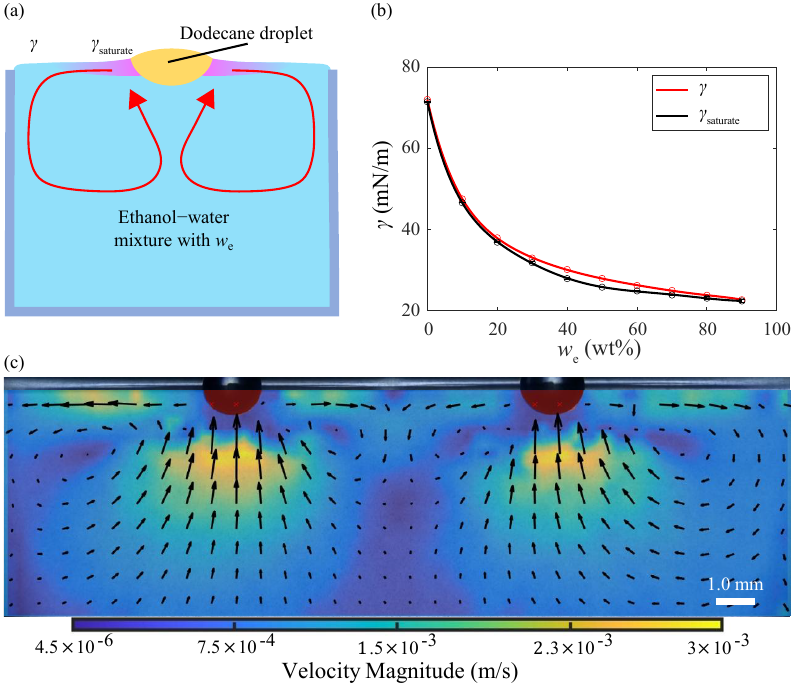}
\caption{(a) A dodecane drop float on the air-liquid interface of an ethanol-water mixture with the ethanol weight fraction $w_\mathrm{e}$.
The gradient of color from pink to blue on the surface of the mixture represents a decreasing concentration of dodecane in the mixture, which results from the dissolution of the drop. The dissolution of dodecane reduces the surface tension of the mixture, thus the surface tension near the drop is lower than in the far field. The consequent Marangoni flow on the pool surface induces a flow inside the container, indicated by the red arrows. 
(b) Surface tension of the dodecane saturated ethanol-water mixture $\gamma_\mathrm{saturate}$ and the ethanol-water mixture free of dodecane $\gamma$ \citep{lide2004crc}. For $\gamma_\mathrm{saturate}$, each point is an average of six measurements and the error bar is the standard deviation. The black and red solid lines are polynomial fittings.  
(c) The flow field for two \SI{3}{\ul} dodecane drops floating on a mixture of $w_\mathrm{e}=\SI{90}{wt\%}$, obtained by PIV measurements. To better visualize the flow field, the two drops are fixed by attaching them to two needles.}
\label{fig:2}
\end{figure*}

\begin{figure*}[t]
\includegraphics[width=1\textwidth]{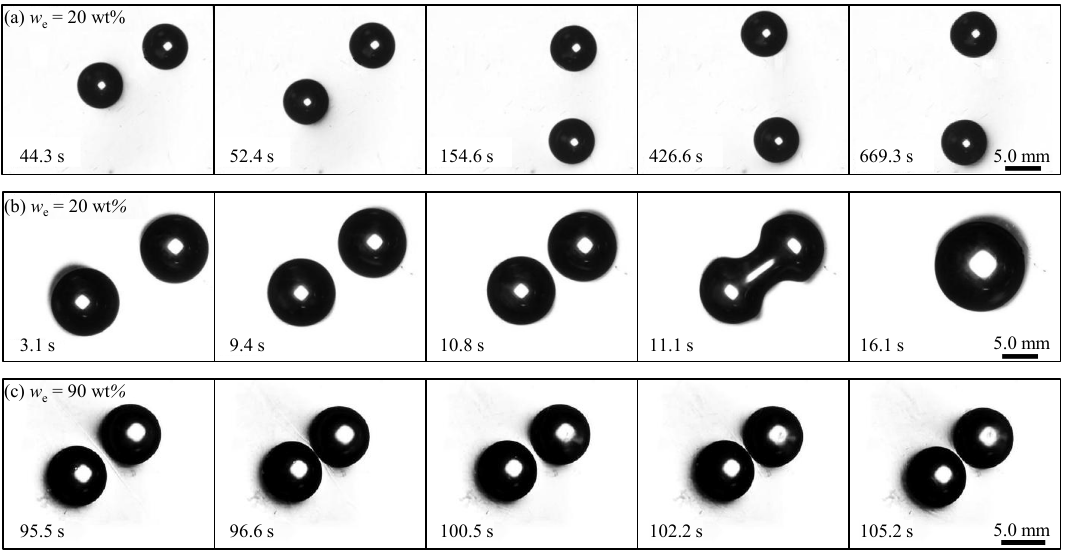}
\captionsetup{singlelinecheck=false, justification=raggedright}
\caption{Three typical behaviors of the two identical drops. (a) Repel. $V=\SI{30}{\micro\liter}$ and $w_\mathrm{e}=\SI{20}{wt\%}$. The drops repel each other. (b) Coalesce. $V=\SI{100}{\micro\liter}$ and $w_\mathrm{e}=\SI{20}{wt\%}$. The drops attract each other and coalesce upon contact. (c) Rebound. $V=\SI{100}{\micro\liter}$ and $w_\mathrm{e}=\SI{90}{wt\%}$. The drops attract each other and rebound upon contact.
} 
\label{fig:behaviors}
\end{figure*}

After the deposition of the drops, dodecane begins to dissolve into the mixture becasue it is partially miscible with the ethanol-water mixture. In a very thin liquid layer close to the drop, the solution becomes saturated very fast \citep{ruschak1972spontaneous}.
The surface tension of the dodecane saturated ethanol-water mixture $\gamma_\mathrm{saturate}$ is found to be always smaller than that of the ethanol-water mixture $\gamma$ free of dodecane (see Fig. \ref{fig:2}(b)), thus the surface tension of the mixture at the edge of the drop is lower than that of the far field, see the sketch in Fig.\ref{fig:2}(a). Consequently, a Marangoni flow is generated on the surface of the pool pointing outwards. When two drops are floating on the pool, the Marangoni flows between the two drops oppose each other, see Fig.\ref{fig:2}(c), which leads to a repulsive force $F_\mathrm{M}$ on the drops. On the other hand, it has been known that a drop floating on a pool will distort the pool's surface \citep{Capillarity_and_Wetting_Phenomena}, either pulling it up or pushing it down. Two drops of the same kind distort the surface in the same way, thus generating an attractive force $F$ on the drops. This is the well known ``Cheerios effect''\citep{Vella2004TheE,Liu2019CapillaritydrivenMO}. The attractive force $F$ and the repulsive force $F_\mathrm{M}$ competes with each other, resulting in different behaviors of the two drops, see Fig. \ref{fig:behaviors} for typical behaviors of the drops. When the ethanol concentration of the pool is low ($w_\mathrm{e}=\SI{20}{wt\%}$) and the drop is relatively small (drop volume $V=\SI{30}{\micro\liter}$), the repulsive force prevails and the distance between the drops increases. We call this behavior ``Repel''. When the ethanol concentration of the pool is kept the same but the drop is larger ($V=\SI{100}{\micro\liter}$), the attractive force prevails and the two drops move closer to each other until finally they coalesce. We call this behavior ``Coalesce''. Interestingly, when the ethanol concentration of the pool is increased to $w_\mathrm{e}=\SI{90}{wt\%}$ while keep the drop volume ($V=\SI{100}{\micro\liter}$), the drops attract each other but they do not coalesce. Instead, they collide with each other and then rebound. We call this behavior ``Rebound''. The drops repeat this attract-rebound process for a long time (\SI{15}{min} for \SI{80}{\micro\liter} drops), until dissolution has made them small enough so that they start to repel each other. 

\begin{figure}[h]
\centering
\includegraphics {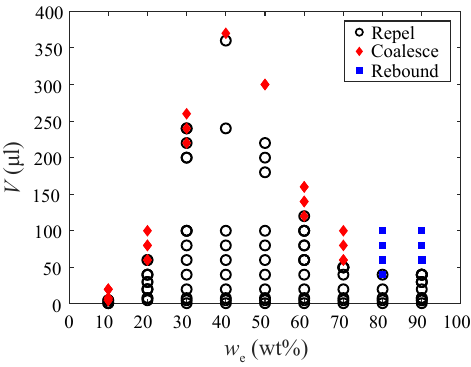}
\captionsetup{singlelinecheck=false, justification=raggedright}
\caption{Phase diagram of the behaviors of the drops. Black circles represent Repel, red diamonds represent Coalesce and blue squares represent Rebound.} 
\label{fig:VvsWe}
\end{figure}

To fully explore the parameter space, the ethanol concentration $w_\mathrm{e}$ of the liquid pool and the drop volume $V$ are varied systematically to perform the experiments and the results are shown in Fig. \ref{fig:VvsWe}. The three different behaviors, ``Repel'', ``Coalesce'' and ``Rebound'' are represented by black circles, red diamonds and blue squares, respectively. It is found that, in general, smaller drops always repel each other. But for each fixed ethanol concentration $w_\mathrm{e}$ of the pool, there is a critical drop volume $V_\mathrm{c}$ above which the drops attract and finally touch with each other. For pool ethanol concentrations equal or smaller than $\SI{70}{wt\%}$, drops will coalesce upon contact, but for larger pool ethanol concentrations, drops will rebound after they collide. We will develop a scaling theory in the next section to explain the transition from repulsion to attraction. Then the transition from coalescence to rebound will be explained in Sec. \ref{From coalescing to rebound}.

\section{Scaling theory for the transition from repulsion to attraction}
\label{Scaling analysis}
To understand the transition from repulsion to attraction, we start by analyzing the two competing forces $F$ and $F_\mathrm{M}$. The attracting force between two floating objects has been studied before and analytical solution exists if the surface deformation caused by the floating particles is small \citep{Nicolson_1949, Vella2004TheE}. However, the solutions were derived for spherical solid particles and does not apply directly for our case where the drop takes the shape of a liquid lens. In the following, we start from the model for spherical solid particles and then modify it to suit our case.

\begin{figure*}[h]
\centering
\includegraphics[width=1\textwidth]{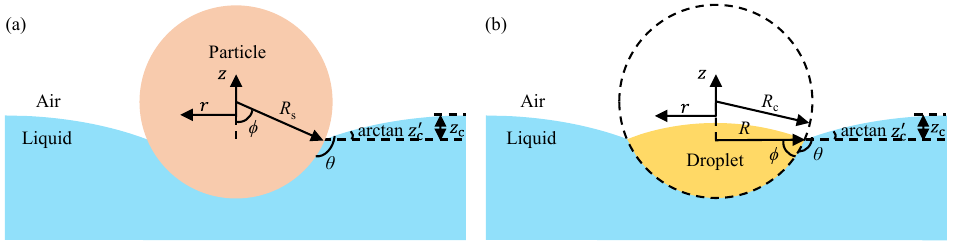}
\caption{Sketch of the distortion of the liquid surface induced by a solid sphere (a) and a liquid drop (b).} 
\label{fig:circle}
\end{figure*}

When a spherical solid particle floats on a liquid pool, it distorts the liquid surface due to gravity, see Fig. \ref{fig:circle}(a). The vertical displacement of the liquid surface is denoted by $z(r)$, where $r$ is the radial distance from the center of the particle and $z = 0$ is the liquid level at infinity.
Thus, the vertical displacement of the contact line is $z_c$. $z_c^\prime$ is the tangent of the liquid surface at the contact line. The polar angle between the south pole of the particle and the position of the contact line is $\phi$. The liquid contact angle is $\theta$. 
Balancing the vertical forces on the particle, we have \citep{Vella2004TheE}
\begin{widetext}
\begin{equation}\label{eq:solid paritcale}
{2\pi\gamma R_s\sin\phi\frac{z_\mathrm{c}{^{\prime}}}{\sqrt{{1+{{z_\mathrm{c}}^{\prime}}}^2}}=\frac{4}{3}\pi\rho_\mathrm{s}gR_s^3-\rho g\pi R_s^3(\frac{z_\mathrm{c}}{R_s}\sin^2\phi+\frac{2}{3}-\cos\phi+\frac{1}{3}\cos^3\phi)},
\end{equation}\end{widetext}
where $R_s$ and $\rho_s$ are the radius and density of the solid particle, respectively. The left-hand-side (LHS) of (\ref{eq:solid paritcale}) is the vertical projection of the surface tension force acting on the particle and the right-hand-side (RHS) is the sum of gravity and buoyancy of the particle. Note that the contact angle $\theta$ does not enter into this equation yet, but it will matter later.

Now let us consider the vertical force balance of a floating drop, which takes the shape of a liquid lens. Since the density difference between dodecane and ethanol-water mixture is relatively small and the drop is not very large, we assume that the upper and lower halves of the liquid lens are two (different) spherical caps. The radius of the lower cap is denoted as $R_c$, see Fig. \ref{fig:circle}(b). Similar to a solid particle, the drop also distorts the pool's surface due to gravity, so that the definition of $z(r)$, $z_c$, $z_c^\prime$, $\phi$ and $\theta$ remain the same. Thus, the vertical projection of the surface tension force acting on the drop takes the same functional form as the LHS of (\ref{eq:solid paritcale}), only replacing $R_s$ with $R_c$. The buoyancy of the drop also takes the same functional form as the second term on the RHS of (\ref{eq:solid paritcale}), again only replacing $R_s$ with $R_c$. The gravity of the floating object, i.e., the first term on the RHS of (\ref{eq:solid paritcale}) should change. But to keep the functional form of (\ref{eq:solid paritcale}), we suggest a pseudo density $\rho_c$ which satisfies
\begin{equation}\label{eq:equivalent density}
\frac{4}{3}\pi \rho_\mathrm{c} R_\mathrm{c}^3=\rho V,
\end{equation}
where $\rho$ and $V$ are the density and volume of the drop, respectively. The pseudo density is thus
\begin{equation}
\rho_\mathrm{c}=\frac{\rho V}{\frac{4}{3}\pi R_\mathrm{c}^3}.
\end{equation}
The vertical force balance on the drop then becomes
\begin{equation}\label{eq:equivalent drops analysis}
2\pi\gamma R_\mathrm{c}\sin\phi\frac{z_\mathrm{c}^{\prime}}{\sqrt{1+{z_\mathrm{c}^{\prime}}^2}}=\frac{4}{3}\pi\rho_\mathrm{c}g R_\mathrm{c}^3 -\rho g\pi R_\mathrm{c}^3(\frac{z_\mathrm{c}}{R_\mathrm{c}}\sin^2\phi+\frac{2}{3}-\cos\phi+\frac{1}{3}\cos^3\phi).
\end{equation}
Notice that $\phi=\pi-\theta+\arctan z^\prime_c$. Following the same procedure as in  \citet{Vella2004TheE}, substitute this into (\ref{eq:equivalent drops analysis}) and keep only terms linear in $z^\prime_c$, we obtaine
\begin{equation}\label{eq:simplify equivalent drops analysis}
z_\mathrm{c}^\prime\sin\phi=Bo(\frac{2D-1}{3}-\frac{1}{2}\cos\theta+\frac{1}{6}\cos^3\theta)\equiv Bo\Sigma,
\end{equation}
where $Bo={R_{\rm{c}}^2}/{{L_\mathrm{c}}^2}$ is the Bond number, $L_\mathrm{c}=\sqrt{{\gamma}/{\rho g}}$ is the capillary length of the mixture, $D={\rho_{\rm{c}}}/{\rho}$ is the density ratio and 
$\Sigma=(\frac{2D-1}{3}-\frac{1}{2}\cos\theta+\frac{1}{6}\cos^3\theta)$. (\ref{eq:simplify equivalent drops analysis}) is accurate to the first order in the Bond number.
Therefore, the net weight of the droplet, i.e., gravity minus buoyancy of the drop, becomes $2\pi\gamma R_\mathrm{c} Bo\Sigma$ in the linearized approximation.

Then the vertical displacement of the pool's surface caused by a floating drop is obtained
\begin{equation}
\label{eq:interfacial deformation}
z(r)=-z_\mathrm{c}^{\prime}L_\mathrm{c}\frac{K_\mathrm{0}(r/L_\mathrm{c})}{K_\mathrm{1}(R_\mathrm{c}\sin\phi/L_\mathrm{c})}\approx-z_\mathrm{c}^{\prime}\sin\phi R_\mathrm{c}K_\mathrm{0}(r/L_{\mathrm{c}}),
\end{equation}
where $K_{\rm{n}}$ is the modified Bessel function of the second kind and of order n.

Having calculated the net weight of a drop at a deformed interface as $2\pi\gamma R_\mathrm{c} Bo\Sigma$ as well as the interfacial deformation caused by the presence of a single drop $z(r)$, we can obtain the energy between two drops with a central distance $L$ away. To leading order in $Bo$, this energy $E(L)$ is obtained by multiplying the net weight of one drop and the vertical displacement caused by the other droplet:
\begin{equation}
\label{eq:surface energy}
E(L)=-2\pi\gamma R_\mathrm{c}^{2}Bo^{2}\Sigma^{2}K_\mathrm{0}(\frac{L}{L_\mathrm{c}}).
\end{equation}
Therefore, the attractive force between the two drops is given as:
\begin{equation}
\label{eq:attraction}
F(L)=-\frac{\mathrm{d}E}{\mathrm{d}L}=-2\pi\gamma R_\mathrm{c} Bo^{\frac{5}{2}}\Sigma^2 K_\mathrm{1}(\frac{L}{L_\mathrm{c}}).
\end{equation}

The repulsive force between the drops is caused by the Marangoni flow, which writes \citep{li2023oilonwater}:
\begin{equation}
\label{eq:repulsion_1}
 F_\mathrm{M}\sim \frac{\Delta\gamma}{L-2R}S,
\end{equation}
where $\Delta\gamma=\gamma-\gamma_\mathrm{saturate}$ and $S$ is the cross-sectional area of the bottom half of the floating drop, which writes $S=R_c^2(\phi-\sin\phi\cos\phi)$. Thus we obtain
\begin{equation}
\label{eq:repulsion}
 F_\mathrm{M}\sim \frac{\Delta\gamma}{L-2R}R_\mathrm{c}^2(\phi-\sin\phi\cos\phi).
\end{equation}

From Eqs.(\ref{eq:attraction}) and (\ref{eq:repulsion}), we find that the attractive force $F\sim R_\mathrm{c}^6$ and the repulsive force $F_\mathrm{M}\sim R_\mathrm{c}^2$. It is then obvious that for small (large) drops, they will repel (attract) each other, which is consistent with the experimental results. 

\begin{figure}[h]
\centering
\includegraphics[scale=1]{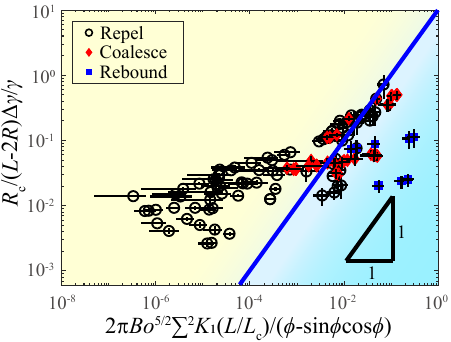}
\caption{According to Eq.\ref{eq:scaling law}, three drops behaviors are plotted in scaling law diagram. 
Black circle represent ``Repel'',  red diamonds stand for ``Coalesce'' and blue squares stand for ``Rebound''. 
The behaviors of drops transform from  ``Repel'' to ``Coalesce'' or ``Rebound'' near the blue critical line.
The error bars are obtained by the law of propagation of uncertainties, see Supplementary Material for more details.
} 
\label{fig:scaling law}
\end{figure}

Balancing the two forces (Eq.(\ref{eq:attraction}) and Eq.(\ref{eq:repulsion})), we obtain the transition criterion from repulsion to attraction:
\begin{equation}
\label{eq:scaling law}
\frac{R_\mathrm{c}}{L-2R}\frac{\Delta\gamma}{\gamma}\sim2\pi Bo^{\frac{5}{2}}\Sigma^2 K_\mathrm{1}\left(\frac {L}{L_\mathrm{c}}\right)\frac{1}{\phi-\sin\phi\cos\phi}.
\end{equation}
Results in Fig. \ref{fig:VvsWe} are replottedwith Eq.(\ref{eq:scaling law}), see Fig. \ref{fig:scaling law}.
To calculate the two terms in Eq.(\ref{eq:scaling law}), $\phi$, $\theta$ (present in $\Sigma$) and $R_\mathrm{c}$ are all obtained from the side view and $R$ and $L$ are measured from the top view. For drops that repel each other, their minimal central distance $L_\mathrm{min}$ is used as $L$. For drops that attract each other (``Coalesce'' and ``Rebound''), their minimal central distance $L_\mathrm{min}= 2R$ so that the LHS of Eq.(\ref{eq:scaling law}) becomes infinitely large. Yet this is not the physical picture, because when the distance between the two drops becomes too small (typically smaller than \SI{100}{\micro\meter}), the dodecane concentration in the pool liquid in between the drops would be relatively large, thus decreasing $\Delta\gamma$. Consequently, the repulsive force does not increase to infinity so that drops can coalesce or rebound. Coming back to the mathematical description of Eq.(\ref{eq:scaling law}), since its LHS becomes infinity for drops that coalesce or rebound, they cannot be plotted in Fig. \ref{fig:scaling law}. But to better show how this scaling works, drops that coalesce or rebound are still plotted by using the $L_\mathrm{min}$ of the critical drop volume $V_\mathrm{c}$. It is found that the scaling law Eq.(\ref{eq:scaling law}) can well separate the repulsive (``Repel'') and attractive behaviors (``Coalesce'' and ``Rebound''). 


\section{From coalescing to rebound}
\label{From coalescing to rebound}
\begin{figure*}[h]
\centering
\includegraphics[width=1\textwidth]{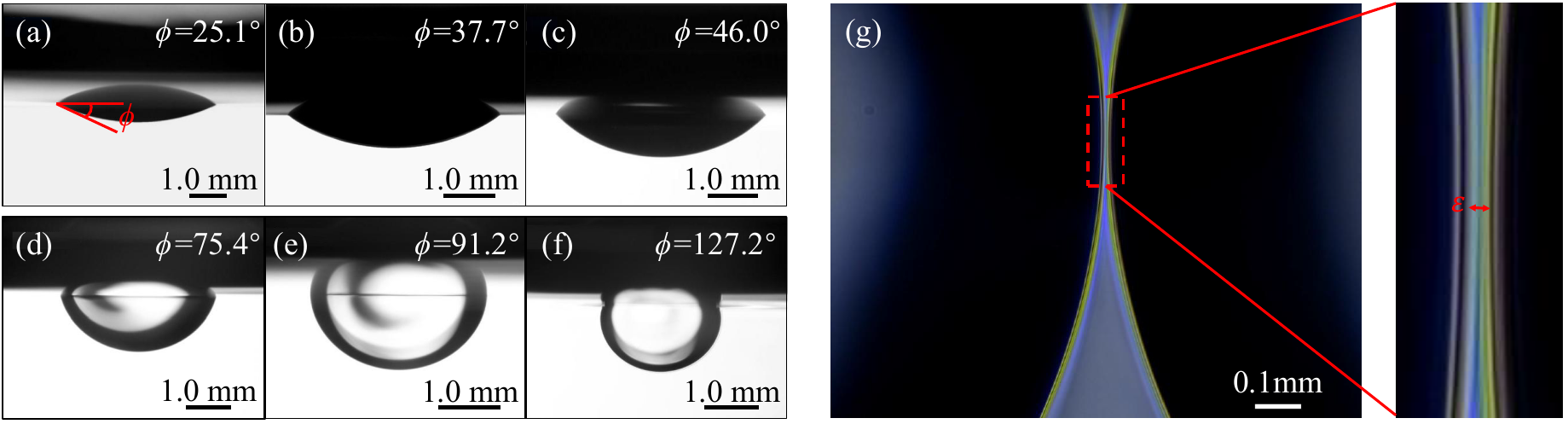}
\captionsetup{singlelinecheck=false, justification=raggedright}
\caption{Shape of the drops ($V=\SI{8}{\ul}$) in the mixture with different ethanol weight fractions $w_\mathrm{e}$: (a) \SI{10}{wt\%}, (b) \SI{20}{wt\%}, (c) \SI{30}{wt\%}, (d) \SI{70}{wt\%}, (e) \SI{80}{wt\%}, (f) \SI{90}{wt\%}. As $w_\mathrm{e}$ increases, angle $\phi$ increases.
(g) A closer look at the thin liquid film between two drops when they collide. The ethanol weight fraction of the pool is $w_\mathrm{e}=\SI{90}{wt\%}$. Thickness of this thin liquid film is denoted by $\varepsilon$.}
\label{fig:side view}
\end{figure*}

As discussed above, for drops that coalesce or rebound, when the drops' distance $L-2R$ becomes very small, the dodecane concentration in the liquid between the two drops also increases, thus $\Delta\gamma$ decreases. In the limit when the two drops touches each other, $\Delta\gamma$ becomes zero so that the repulsive force $F_\mathrm{M}$ vanishes. Consequently, there must be another repulsive mechanism that leads to the transition from coalescing to rebound. The fact that the drops could rebound several times (see Fig. \ref{fig:behaviors}) indicates that this repulsive mechanism is only effective at a very short length scale, because the drops attract each other again once they are slightly apart, see the 3rd snapshot in Fig. \ref{fig:behaviors}(c). We suspect that under these conditions ($w_\mathrm{e}\geq\SI{80}{wt\%}$), a lubrication film forms when the drops collide with each other. The shapes of the lower halves of the drops at different pool ethanol concentrations were observed and the half central angles of the lower halves $\phi$ were measured, see Fig.\ref{fig:side view}. It is found that for pool ethanol concentration $w_\mathrm{e}\geq\SI{80}{wt\%}$, $\phi$ becomes larger than $\SI{90}{\degree}$. This would facilitate the formation of a lubrication film, which could provide a strong enough repulsive force that leads to the rebound of the drops. 

To further confirm this, we take a closer look at the gap between two $\SI{100}{\micro\liter}$ drops right before they rebound (in a pool of ethanol concentration $w_\mathrm{e}=\SI{90}{wt\%}$), see Fig.\ref{fig:side view}(g). Indeed, a small gap exists in between the drops and liquid in the gap flows downwards due to the Marangoni flow. Denoting the thickness of this thin liquid film as $\varepsilon$, we know that the pressure difference that such a film could withstand is $\Delta p\sim O(1/\varepsilon^{\alpha})$ \citep{Levitation_article}.
When the drop is immiscible with the film liquid, $\alpha=2$. If the drop dissolves into the thin film, then $\alpha>2$. On the other hand, from Eq.\ref{eq:attraction} we know that the attractive force between the two drops is proportional to $K_1(L/L_\mathrm{c})$. Since $K_1(L/L_\mathrm{c})\approx L_\mathrm{c}/L$ when $L$ is very small, the attractive pressure $p_a$ between the two drops caused by the Cheerios effect is $p_a\sim O(1/L)$. Because $\varepsilon<L$ and $\alpha\geq2$, the lubrication induced repulsive force will be larger that the attractive force as long as the two drops are close enough. Consequently, the drops will rebound once they collide. 

\section{Conclusions}
\label{CONCLUSIONs}
In summary, the interactions of two drops floating on partially miscible ethanol-water mixtures are investigated experimentally and theoretically.
The drop volume $V$ and the ethanol weight fraction $w_\mathrm{e}$ of the mixture are systematically varied to explore the parameter space. Three typical behaviors are discovered: (i) Repel. The drops repel each other and their central distance increases. (ii) Coalesce. The drops attract each other and coalesces upon contact. (iii) Rebound. The drops attract each other but rebound upon contact. The repulsive or attractive behavior between the drops are controlled by the competition between two forces: an attractive force due to the Cheerios effect and a repulsive force due to the opposing Marangoni flows caused by drop dissolution. A scaling theory was developed to distinguish the repulsive and attractive behaviors, which fits well with the experimental results. For drops that attract each other, the drops' shape determines whether they will coalesce or rebound. When the half central angle of the drop's lower half is obtuse, which happens for pool ethanol concentration equal or greater than $\SI{80}{wt\%}$, a lubrication film forms between the two drops upon contact, preventing them from coalescing, which eventually leads to the rebound of the drops. If this angle is acute, the drops would coalesce upon contact, which is the case for pool ethanol concentration smaller than $\SI{80}{wt\%}$. 

In our research, the drop dissolution decreases the surface tension of the liquid pool, generating an outward Marangoni flow and consequently a repulsive force between drops. It might be interesting to see what will happen if the dissolution increases the surface tension of the pool.

\section{Data Availability}
\label{Data Availability}
The data supporting the findings of this study are available from the corresponding author upon reasonable request.

\section{Author Contributions}
\label{Author Contributions}
Yuan Gao: Investigation, Data curation, Visualization, Writing original draft. Yanshen Li: Conceptualization, Methodology, Supervision, Writing–review and editing.

\section{Acknowledgements}
We acknowledge the financial support from the National Natural Science Foundation of China under grant No. 12272376.


\appendix

\bibliographystyle{elsarticle-harv-URL-DOI.bst}
\bibliography{References}

\end{document}